\documentclass[amsmath,amssymb,aps,prl,reprint,superscriptaddress]{revtex4-1}

\usepackage{tensor}
\usepackage{graphicx}
\usepackage{amsmath, amssymb, amsfonts, mathtools}
\usepackage{ulem}
\usepackage{float}
\usepackage{color}
\usepackage{natbib}

\newcommand\apjl{ApJ}

\newcommand\aap{A\&A}
\newcommand\mnras{MNRAS}

\newcommand \rhoe {\ensuremath{ \rho_{\rm e} }}
\newcommand \divD {\nabla \cdot \vec{D}}

\newcommand \curlE {\nabla \times \vec{E}}
\newcommand \curlH {\nabla \times \vec{H}}
\newcommand \E {\vec{E}}
\newcommand \B {\vec{B}}
\newcommand \D {\vec{D}}
\renewcommand \H {\vec{H}}
\newcommand \J {\vec{J}}
\renewcommand \j {\vec{j}}
\newcommand \vbeta {\vec{\beta}}
\newcommand \rg {\ensuremath{r_{\rm g}}}
\newcommand \rH {\ensuremath{r_{\rm H}}}
\newcommand \rlarmor {\ensuremath{r_{\rm L,0}}}
\newcommand \OmegaH {\ensuremath{\Omega_{\rm H}}}
\newcommand \OmegaB {\ensuremath{\Omega_{\rm B_0}}}
\newcommand \skin {\ensuremath{\delta_{0}}}

\newcommand \dd { \partial }
\newcommand \DD { {\rm d} }
\renewcommand{\vec}[1]{\boldsymbol{#1}} 

\begin{document}

\title{First-Principles Plasma Simulations of Black-Hole Jet Launching}

\author{Kyle Parfrey}
\email{kparfrey@lbl.gov}
\altaffiliation{Einstein Fellow}
\affiliation{Lawrence Berkeley National Laboratory, 1 Cyclotron Road, Berkeley, CA 94720, USA}
\affiliation{Department of Astronomy and Theoretical Astrophysics Center, UC Berkeley, Berkeley, CA 94720, USA}
\affiliation{NASA Goddard Space Flight Center, Mail Code 661, Greenbelt, MD 20771, USA}

\author{Alexander Philippov}
\altaffiliation{Einstein Fellow}
\affiliation{Department of Astronomy and Theoretical Astrophysics Center, UC Berkeley, Berkeley, CA 94720, USA}
\affiliation{Center for Computational Astrophysics, Flatiron Institute, 162 Fifth Avenue, New York, NY 10010, USA}

\author{Beno\^it Cerutti}
\affiliation{Univ.\ Grenoble Alpes, CNRS, IPAG, 38000 Grenoble, France}

\begin{abstract}
Black holes drive powerful plasma jets to relativistic velocities. This plasma should be collisionless, and self-consistently supplied by pair creation near the horizon. We present general-relativistic collisionless plasma simulations of Kerr-black-hole magnetospheres which begin from vacuum, inject e$^\pm$ pairs based on local unscreened electric fields, and reach steady states with electromagnetically powered Blandford-Znajek jets and persistent current sheets. Particles with negative energy-at-infinity are a general feature, and can contribute significantly to black-hole rotational-energy extraction in a variant of the Penrose process. The generated plasma distribution depends on the pair-creation environment, and we describe two distinct realizations of the force-free electrodynamic solution. This sensitivity suggests that plasma kinetics will be useful in interpreting future horizon-resolving submillimeter and infrared observations.
\end{abstract}

\maketitle

The relativistic jets of plasma emanating from active galactic nuclei and X-ray binary systems are widely thought to be driven by magnetic fields threading a rotating black hole, known as the Blandford-Znajek mechanism \cite{Blandford:1977aa}. This process is generally studied using magnetohydrodynamics (MHD), a fluid approximation for the plasma. While MHD has facilitated significant progress in understanding black-hole accretion and jet production \cite{Koide:1998aa,De-Villiers:2003aa,McKinney:2004aa,Komissarov:2004aa,Tchekhovskoy:2011,Foucart:2016aa}, it suffers several shortcomings which limit its descriptive power for this problem; for example, the pair-creation process which supplies the jet with electron-positron plasma \cite{Beskin:1992ab,Hirotani:1998aa,Vincent:2010aa,Broderick:2015aa,Ptitsyna:2016aa,Levinson:2017aa,Ford:2017aa} cannot be captured within MHD, which therefore cannot predict the jet's mass loading.

Furthermore, the jets have low densities and hence particles have large mean free paths between two-particle collisions. The plasma is effectively collisionless, as is that in many low-luminosity black-hole accretion flows, including those of Sgr A* and M87 \cite{Narayan:1995aa,Quataert:2003aa}, the targets of ongoing campaigns to resolve horizon-scale structures by the Event Horizon Telescope (EHT) \cite{Doeleman:2008aa,Doeleman:2012aa} and GRAVITY \cite{Hamaus:2009aa,Gravity-Collaboration:2017aa}.
Collisionless plasmas support complex behavior that can only be reflected by the full system of plasma kinetics, which can self-consistently describe the non-ideal unscreened electric field, pair creation, particle acceleration, and the emission of observable radiation.

Recently there has been progress on local, one-dimensional simulations of the electrostatic physics of black holes' vacuum gaps \cite{Daniel:1997aa,Levinson:2018aa,Chen:2018aa}, which must be embedded in an assumed field and current configuration. Global models are required to self-consistently include the feedback of the plasma on the magnetosphere, and have been used to study the earth's magnetosphere \cite{Lin:2005aa,von-Alfthan:2014aa} and those of radio pulsars \cite{Chen:2014aa,Philippov:2014aa}. Here we present the first global, fully general-relativistic, multi-dimensional kinetic simulations of black-hole magnetospheres.

We solve the kinetic system using the particle-in-cell approach with a code based on \textsc{zeltron} \cite{Cerutti:2013aa}, and express the equations for the particles and fields using the 3+1 formalism. We use geometrized units with $G=M=c=1$, where $M$ is the black hole's mass; lengths are given in units of $\rg=GM/c^2$, and times are in $\rg/c$. It is useful to consider the local fiducial observer (FIDO) which is normal to spatial hypersurfaces, having 4-velocity $n_\mu = (-\alpha,\vec{0})$, $n^\mu = (1, -\vbeta)/\alpha$.


The fields are evolved with Maxwell's equations,
\begin{align}
\label{eq:FaradaysLaw}
    \dd_t \B &= - \curlE , \\
    \dd_t \D &= \curlH - 4\pi\J ,
\label{eq:AmperesLaw}
\end{align}
where $\B$ and $\D$ are the magnetic and electric fields measured by FIDOs, and the auxiliary fields are $\H=\alpha\B-\vbeta\times\D$ and $\E=\alpha\D+\vbeta\times\B$ \cite{Komissarov:2004ab}. The current measured by FIDOs is $\j = (\J + \rhoe \vbeta)/\alpha$, where $4\pi\rhoe = \divD$, and the 4-current is $I^\mu = (\rhoe, \J)/\alpha$. 

The fields are staggered in space on a Yee mesh and offset in time by half a step. They are evolved with a new leapfrog integrator, which uses trapezoidal leapfrogs for the $\vbeta\times\B$ and $\vbeta\times\D$ terms inside the curls.


The particles' equations of motion can be written as the Hamiltonian pair
\begin{align}
\label{eq:posEOM}
\frac{\DD x^i}{\DD t}&=v^i=\frac{\alpha}{\Gamma}\gamma^{ij}u_j-\beta^i,\\
\frac{\DD u_i}{\DD t}&=-\Gamma\dd_i\alpha + u_j\dd_i\beta^j 
        -\frac{\alpha}{2\Gamma}\dd_i(\gamma^{lm}) u_l u_m 
        +\frac{\alpha}{m}\mathcal{L}_{i},
\label{eq:momEOM}
\end{align}
where $\gamma^{ij}$ is the inverse of the spatial 3-metric, $u_{\mu} = (u_t,u_i)$ is the particle's 4-velocity, and the FIDO measures the particle's Lorentz factor, Lorentz force, and 3-velocity to be $\Gamma = \sqrt{1 + \gamma^{ij}u_iu_j}$, $\vec{\mathcal{L}} = e\,(\D+\vec{V}\times\B)$, and $\vec{V}=(\vec{v}+\vbeta)/\alpha$ respectively. The particles have mass $m$ and charge $e$. They supply the current $\J$ required by Eqn.~(\ref{eq:AmperesLaw}), with the contribution from each particle being proportional to $e \vec{v}$ and assigned to grid locations with the volume-weighting technique; this closes the system. Gauss's law is enforced by the frequent use of a Poisson solver to correct the $\D$ field.

We evolve the particles with a time-symmetric Strang splitting of Eqns.~(\ref{eq:posEOM}) and (\ref{eq:momEOM}). First the momentum is pushed forward by half a time step with the Lorentz force term alone, using the Boris algorithm in the FIDO's local frame by means of a tetrad basis. Then the position equation, and the gravitational and coordinate terms in the momentum equation, are evolved together for a full time step, using an iterative symplectic integrator. Finally the Lorentz force again acts on the momentum for another half time step. This scheme conserves energy in the absence of electric fields, and is relatively computationally cheap. The numerical methods developed for the field and particle evolution will be described in detail in a future paper.


\begin{figure}[t]
  \begin{center}
    \includegraphics[width=8.6cm, trim = 3.0mm 2.0mm 2.5mm 2.0mm, clip]{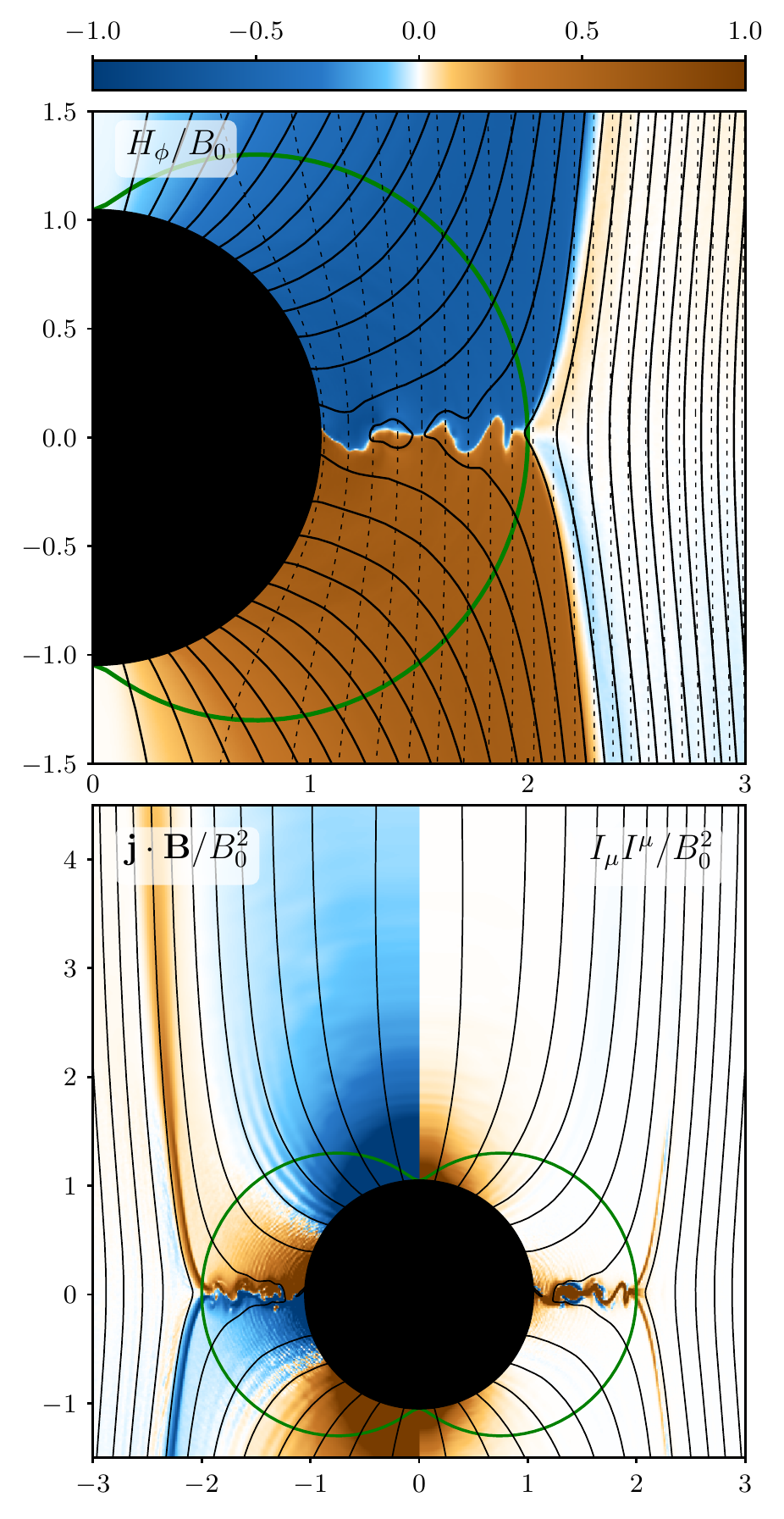}
  \end{center}
  \vspace{-5mm}
  \caption{ \label{fig:Hphi} Toroidal magnetic field, field-aligned current, and 4-current norm for the high-plasma-supply scenario in the steady state. The ergosphere boundary is shown in green, and magnetic flux surfaces are in black; dashed lines indicate the same flux surfaces in the initial Wald state.}
  \vspace{-6mm}
\end{figure}

Our initial field configuration is Wald's stationary vacuum solution for a rotating black hole immersed in an asymptotically uniform magnetic field, aligned with the hole's angular-momentum vector, which includes the electric field generated by spacetime rotation \cite{Wald:1974aa}. There are no particles in the initial state. We use the Kerr metric with spin parameter $a = 0.999$ and the Kerr-Schild spacetime foliation in spherical coordinates ($r,\theta,\phi$). Here we focus on two high-resolution simulations; we also performed several runs at lower resolution to infer the dependence on various parameters.

We set the field strength at infinity to $B_0=10^3\,m/|e|$, so moderately relativistic particles initially have Larmor radii $\rlarmor\sim10^{-3}$ and gyro frequencies $\OmegaB=10^3$. This provides a reference scale for many quantities, such as the Goldreich-Julian number density $n_0=\Omega_{\rm H}B_0/4\pi e$, where $\OmegaH=a/(\rH^2+a^2)$ is the angular velocity of the horizon at $r=\rH$, and the magnetization $\sigma_0=B_0^2/4\pi n_0m=\OmegaB/\OmegaH\approx2000$. These scales imply the astrophysically relevant ordering $\rlarmor\ll\skin\ll\rg$, where $\skin=\sqrt{\sigma_0}\,\rlarmor$ is the skin depth.

The axisymmetric computational domain covers $0.985\,\rH\leq r\leq8$ and $0\leq\theta\leq\pi$.  The grid consists of $N_r\times N_\theta=1280\times 1280$ cells, equally spaced in $\log r$ and $\cos\theta$, which concentrates resolution toward the horizon and the equator. The simulations have duration $\Delta t=50$. Waves and particles are absorbed in a layer at the outer boundary \cite{Cerutti:2015aa}. The inner boundary lies inside the horizon and all equations are solved there without modification.

\begin{figure*}
  \begin{center}
    \includegraphics[width=\textwidth, trim = 2.0mm 1.0mm 1.0mm 2.0mm, clip]{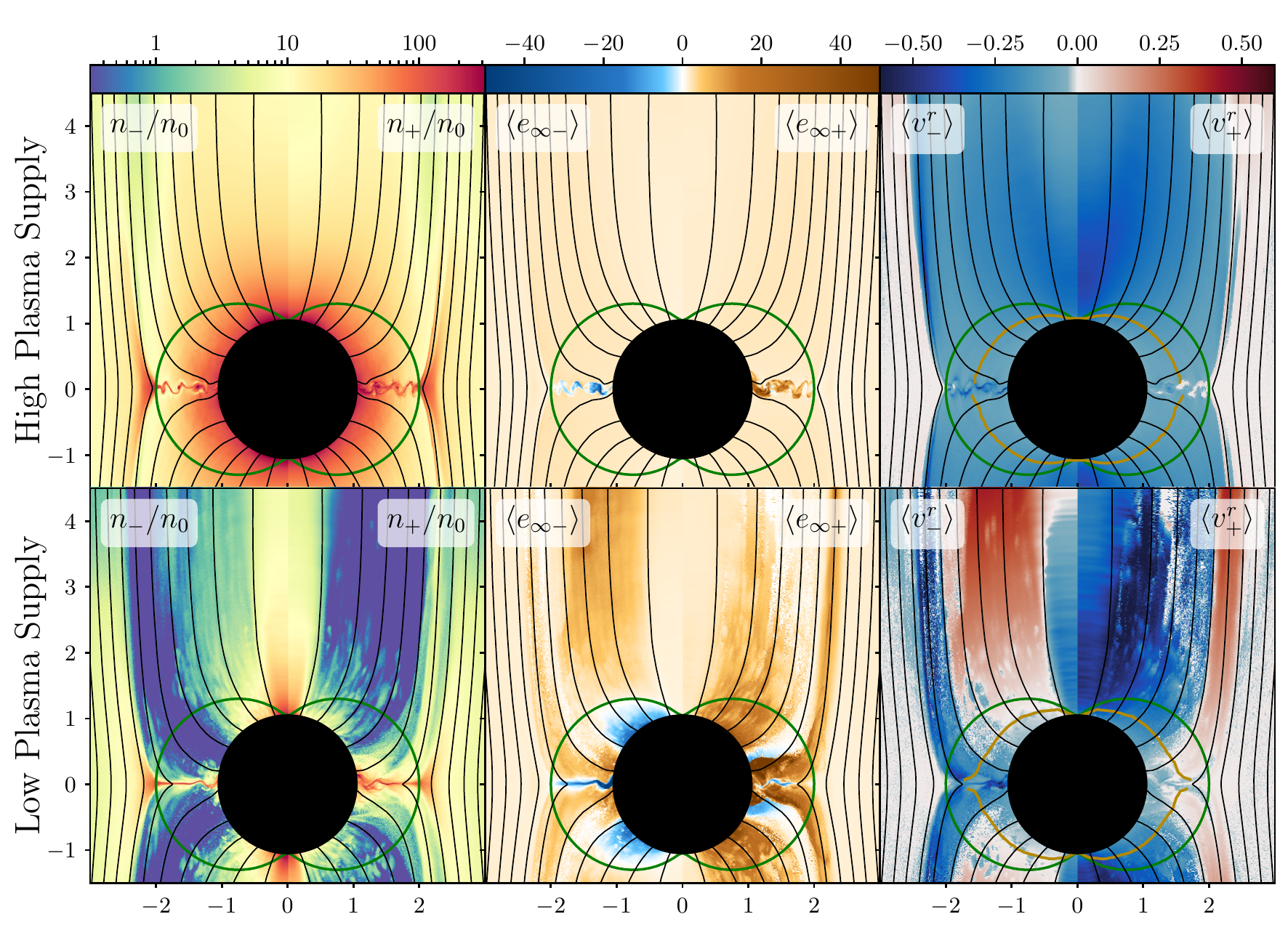}
  \end{center}
  \vspace{-5mm}
  \caption{ \label{fig:nuv} Steady-state FIDO-frame density, and average particle energy-at-infinity and radial velocity (Eqn.~\ref{eq:posEOM}), for electrons ($-$) and positrons ($+$) in the two plasma-supply scenarios. Note the regions where the average particle energy is negative --- the black hole's rotational energy decreases when these particles cross the horizon. Gold lines indicate the inner light surface.} 
  \vspace{-5mm}
\end{figure*}

Plasma is introduced throughout the simulation in the volume $\rH<r<6$. We defer a realistic treatment of pair-creation physics to future work, and instead use a simple prescription which allows us to specify how precisely the force-free $\D\cdot\B=0$ condition is satisfied \cite{Belyaev:2015ab}. In each cell, at each time step, an electron-positron pair is injected, with each particle conferring an effective FIDO-measured density of 
\begin{equation}
    \delta n_{\rm inject} = \frac{\mathcal{R}}{4\pi e} \frac{|\D\cdot\B|}{B},
    \label{eq:ninject}
\end{equation}
provided that $|\D\cdot\B|/B^2$ is greater than a threshold $\epsilon_{D\cdot B}$, and that the non-relativistic magnetization $\sigma>\sigma_0/20$. We set $\mathcal{R}=0.5$ and create two scenarios, motivated by the range of pair-creation environments around astrophysical black holes: a ``high plasma supply'' scenario with a small pair-creation threshold, $\epsilon_{D\cdot B}=10^{-3}$, and one with ``low plasma supply'' where $\epsilon_{D\cdot B}=10^{-2}$. These different pair-injection thresholds lead to two distinct states of the system. The particles are injected with velocities randomly drawn from a relativistic Maxwellian of temperature $k_{\rm B}T=0.5\,m$.


The evolution in the two plasma-supply scenarios is similar in many respects. The Wald solution for $a\sim1$ has large parallel electric fields induced by spacetime rotation, $|\D\cdot\B|\sim B^2$, and so when the simulation begins the magnetosphere rapidly fills with plasma, following Eqn.~(\ref{eq:ninject}). This plasma produces currents which drive the system away from the vacuum steady state. The magnetic field lines, which were originally nearly perfectly excluded from the horizon, now bend back toward the black hole and penetrate the horizon. The bending is only severe inside the ergosphere, which extends to $r=2$ on the equator. Plasma falls along the field lines toward the hole, and accumulates at the equator on those ergospheric field lines which do not yet cross the horizon. An equatorial current sheet forms, initially at the horizon and rapidly extending to the ergosphere boundary.

By $t\sim20$ almost all field lines which enter the ergosphere also cross the horizon. The thin current sheet is then disrupted by the drift-kink instability, which begins at the horizon and moves outward. Magnetic reconnection occurs across the sheet, leading to the formation of isolated plasmoids, which move inward and through the horizon. The entire magnetosphere enters a long-term quasi-equilibrium by $t\sim40$. All figures show the two simulations at the same two steady-state reference times, $t_{\rm ref}^{\rm high}\sim40$ and $t_{\rm ref}^{\rm low}\sim48$.


In this approximate steady state, the toroidal magnetic field is large in the jet, which consists of those field lines which enter the ergosphere, and very small outside it (Fig.~\ref{fig:Hphi}, top). There is a strong current layer along the jet boundary as well as volume currents of both directions inside the jet; the current is highly spacelike at the poles and along the equatorial current sheet and the boundary current layer, requiring the presence of both particle species, and nearly null elsewhere (Fig.~\ref{fig:Hphi}, bottom). In the high-supply case, the FIDO-measured density of both species is well above the reference value everywhere, $n\sim10$--$100\,n_0$ (Fig.~\ref{fig:nuv}, upper left); there are $\sim10^3$ particles per species per cell, and $\sim3\times10^9$ particles in total.

The low-supply simulation initially evolves similarly, but starting at $t\sim15$ the density of both species inside the jet begins to drop, and the electrons begin to flow away from the hole. This counter-streaming allows the limited charges to carry the current required by the global magnetosphere, which is similar to that shown in Fig.~\ref{fig:Hphi}, though with more-diffuse high-current structures. This is to be expected, as in both scenarios the deviations from the force-free electrodynamic solution are small. Now the densities are generally much lower and the jet is largely charge-separated, with electrons in the polar region and positrons in a thick layer along the jet boundary (Fig.~\ref{fig:nuv}, lower left).

\begin{figure}
  \begin{center}
    \includegraphics[width=8.6cm, trim = 1.0mm 1.0mm 0.0mm 0.0mm, clip]{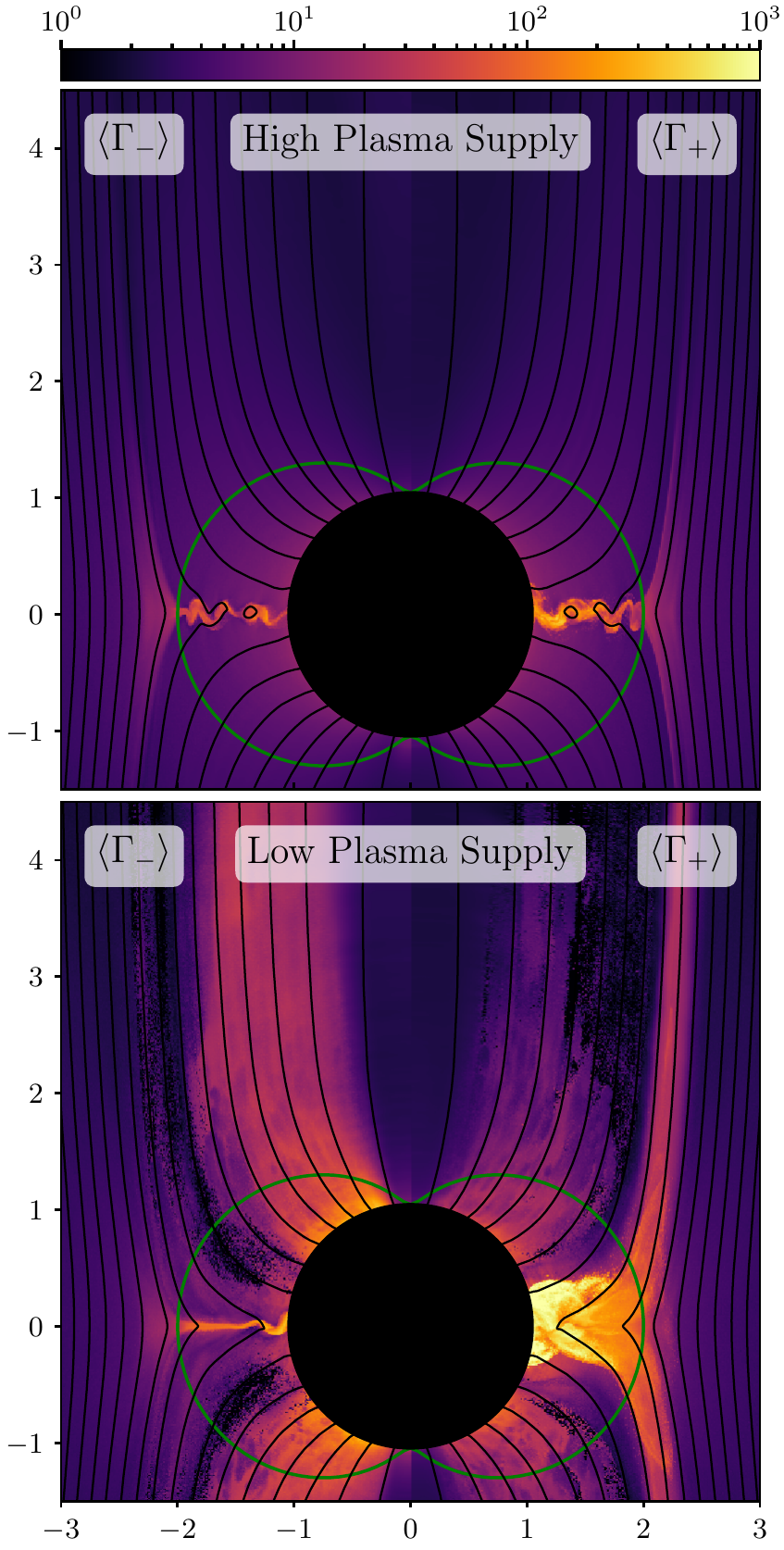}
  \end{center}
    \vspace{-5mm}
  \caption{\label{fig:gamma} Average FIDO-measured Lorentz factors in the two steady states; the full potential corresponds to $\Gamma_{\rm max}\sim10^3$.} 
    \vspace{-5mm}
\end{figure}


Both simulations contain particles which have negative energy-at-infinity $e_\infty=-u_t$, due to the action of the Lorentz force (Fig~\ref{fig:nuv}, center). Penrose has proposed the ingestion of these particles as a mechanism to extract a black hole's rotational energy \cite{Penrose:1969aa}. MHD simulations have shown bulk negative-energy regions in transient behavior \cite{Koide:2002aa,Koide:2003aa} but not in the steady state \cite{Komissarov:2005aa}. 
Our kinetic simulations demonstrate that non-ideal electric fields, from reconnection and charge starvation, continue to push particles onto negative-energy trajectories.
In both scenarios, electrons are given negative energies in the current sheet, where the average electron velocity is toward the black hole. They cross the horizon and extract the hole's energy and angular momentum. The low-supply run also shows negative-energy electrons in the electron-dominated part of the jet, with the $\langle e_{\infty-}\rangle<0$ region extending up to the ergosphere boundary, beyond which this effect is impossible; angled brackets imply averaging over the distribution within one cell.

Some of these polar negative-energy electrons also flow into the hole. The electrons have a velocity-separation surface in the jet coincident with the inner light surface, at which corotation with the field lines at $\Omega_{\rm F}=-E_\theta/\sqrt{\gamma}B^r\approx\OmegaH/2$ and fixed $(r,\theta)$ is a null world-line (Fig.~\ref{fig:nuv}, lower right). Electrons have $\langle v^r_-\rangle<0$ on field lines close to the polar axis. The positrons also show radial-velocity separation at the light surface, in the jet-boundary region where their average velocity at large radii is positive. In contrast, the high-supply scenario shows negative radial velocity for both species throughout the jet, with the current supported by comparatively small velocity differences.


Particles are accelerated to high energies in the equatorial current sheet, with many reaching the approximate limiting Lorentz factor implied by the total potential drop, $\Gamma_{\rm max}\sim a\,\OmegaB\sim10^3$; see locally averaged values in Fig.~\ref{fig:gamma}. In both scenarios the positrons have higher FIDO-frame Lorentz factor in the current sheet. The particles are roughly an order of magnitude more energetic in the low-supply case, with accelerated electrons (positrons) having positive $\langle v^r\rangle$ in the jet (jet-boundary current layer). In this scenario the region of positrons accelerated to $\sim\Gamma_{\rm max}$ is much thicker than the current sheet itself. In all cases the particle energies are low near the polar axis.


\begin{figure}
  \begin{center}
    \includegraphics[width=8.6cm, trim = 0.0mm 0.0mm 0.0mm 0.0mm, clip]{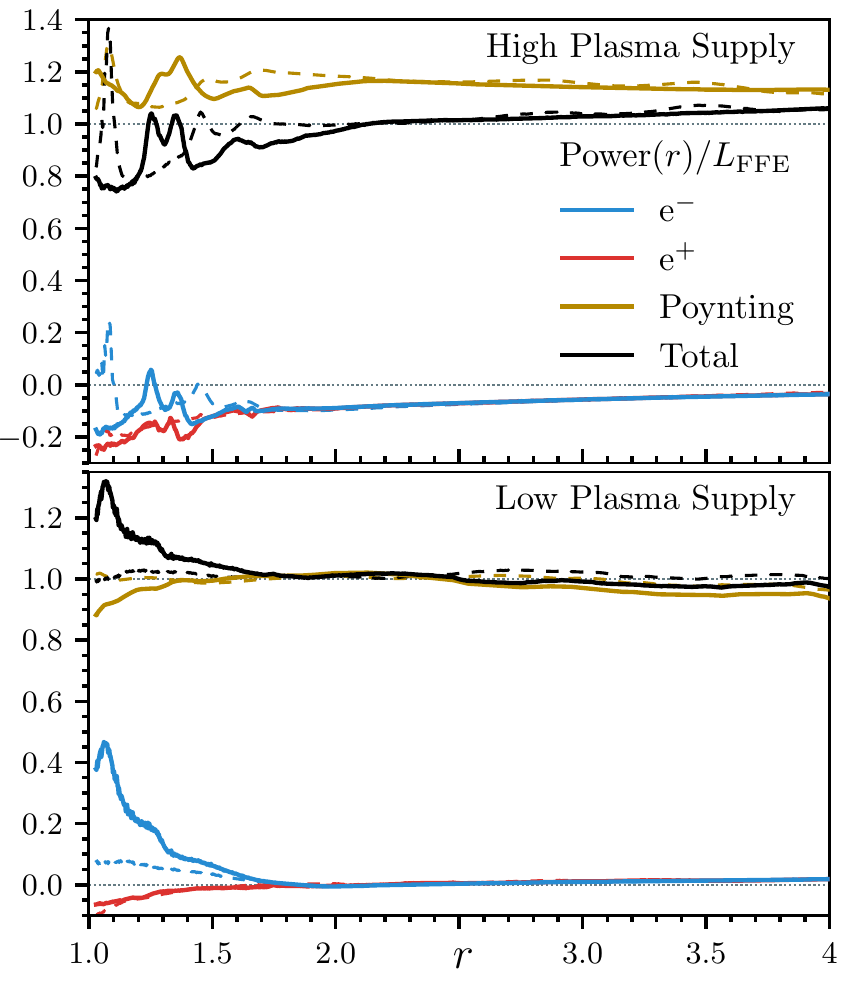}
  \end{center}
  \vspace{-5mm}
  \caption{ \label{fig:fluxes} Flux of energy-at-infinity through spherical shells in units of the force-free value, for two steady-state epochs (solid curves: $t_{\rm ref}$; dashed curves: $t=50$). Positive values at $\rH\sim1$ imply extraction of the black hole's rotational energy.} 
  \vspace{-5mm}
\end{figure}

The total flux of conserved energy-at-infinity passing through spherical shells, $\int T^r_t\alpha\sqrt{\gamma}\,\DD\theta\DD\phi$ where $T^{\mu\nu}$ is the total energy-momentum tensor of the plasma and the fields, is roughly constant on average and of comparable magnitude to the corresponding force-free solution, $L_{\rm FFE}\approx0.2\,B_0^2$ (Fig.~\ref{fig:fluxes}), as found with the \textsc{phaedra} code \cite{Parfrey:2012ab,Parfrey:2015aa}. Current-sheet instabilities produce fluctuations inside the ergosphere at up to the 50\% level, with variations in the high-supply scenario being generally larger. 

Far from the horizon, the energy flux in the particles, given by $T^r_{t,\pm}\alpha=\langle e_{\infty\pm}\,v^{r}_\pm\rangle\,n_\pm$, is small and the jet power is almost entirely transmitted as Poynting flux, with $T^r_{t,{\rm EM}}\alpha=(\E\times\H)^r/4\pi$. Inside the ergosphere the energy flux carried by the particles can be large. 

In the denser high-supply simulation, the large inward flux of positive particle energy from both species in the jet usually, though not always, exceeds the energy-extracting contribution from inflowing negative-energy electrons in the current sheet. In the low-supply scenario, the positron energy-flux contribution is usually small and negative, while that from the electrons is almost invariably positive and carries up to $\sim0.5\,L_{\rm FFE}$. This demonstrates that the ingoing negative-energy ``Penrose'' particles can become the dominant component, making the total particle population a net contributor to black-hole rotational-energy extraction.


In other simulations, we inject particles isotropically in the frame of the Boyer-Lindquist normal observer, leading to nearly identical results, including for the detailed velocity structure shown in Fig.~\ref{fig:nuv}.
Simulations with lower $B_0$, and hence lower $\sigma_0$, confirm that as magnetization increases the fraction of the energy flux carried by infalling positive-energy particles declines. At high $B_0$ we expect the energy-extracting current-sheet electrons to always dominate the particle energy flux.

We performed simulations in which $\alpha$, $\vbeta$, and $\gamma^{ij}$ were set to their flat-spacetime values inside the derivative terms in the particle momentum equation, Eqn.~(\ref{eq:momEOM}), which removes all of the effective gravitational forces. Now the particle evolution does not conserve energy and momentum, and accelerated high-$\Gamma$ particles, whose large Larmor radii allow them to experience the incorrect gradient terms, drive unphysical currents which eventually destroy the solution. Additionally, the high-density region near the poles in the low-supply state, coincident with the $\langle v^r_-\rangle<0$ polar region (Fig.~\ref{fig:nuv}), does not exist without gravity; rather, both species have lower densities, and electrons have positive radial velocities. We speculate that, near the pole, the nearly field-aligned gravitational forces interfere with charge redistribution by the parallel electric field, leading to less efficient screening, more particle injection, and a polar region resembling the high-supply state.


We have described the first direct plasma-kinetic simulations of the Blandford-Znajek process, in which a plasma-filled magnetosphere mediates the extraction of a black hole's rotational energy and the launching of a relativistic jet. We show that the plasma distribution is sensitive to the pair-supply mechanism, and describe two distinct states, both electrodynamically similar to the force-free solution, which would lead to highly dissimilar observable emission. When a particle species has a velocity-separation surface in the jet, the jet's currents are partly carried by an ergospheric population with negative energy-at-infinity, implying a supporting role for the Penrose process in general Blandford-Znajek jets. Our simulations also have a current sheet at the equator, where the Penrose effect can be responsible for a large fraction of the total energy flux from the black hole. Future simulations will include a more realistic treatment of the pair-creation physics, allowing us to model the accelerating electrostatic gaps, and the resulting photon emission, in the context of a self-consistent global magnetosphere, enabling a rigorous interpretation of the EHT and GRAVITY observations.

\vspace{2mm}

The authors would like to thank E.\ Quataert, M.\ Medvedev, A.\ Levinson, V.\ Beskin, J.\ Mortier, and Z.\ Meliani for useful discussions. KP and AP were supported by NASA through Einstein Postdoctoral Fellowships, with grant numbers PF5-160142 and PF7-180165 respectively. BC was supported by CNES and Labex OSUG@2020 (ANR10 LABX56). Resources supporting this work were provided by the NASA High-End Computing Program through the NASA Advanced Supercomputing Division at Ames Research Center, and TGCC and CINES under the allocation A0030407669 made by GENCI. Research at the Flatiron Institute is supported by the Simons Foundation.

\end{document}